\documentclass[preprint]{revtex4-2}
\usepackage{amssymb}
\usepackage{amsmath}
\usepackage{bm}

\usepackage{epsfig}
\usepackage{tabularx}
\usepackage{array}
\usepackage{booktabs}

\usepackage{color}
\usepackage{braket}

\usepackage{hyperref}

\begin{document}

\title{Nonadiabatic H-Atom Scattering Channels on Ge(111) Elucidated by the Hierarchical Equations of Motion}
\author{Xiaohan Dan}
\affiliation{Department of Chemistry, Yale University, New Haven, CT 06520,
USA}
\author{Zhuoran Long}
\affiliation{Department of Chemistry, Yale University, New Haven, CT 06520, USA}
\author{Tianyin Qiu}
\affiliation{Department of Chemistry, Yale University, New Haven, CT 06520, USA}
\author{Jan Paul Menzel}
\affiliation{Department of Chemistry, Yale University, New Haven, CT 06520, USA}

\author{Qiang Shi}\email{qshi@iccas.ac.cn}
\affiliation{Beijing National Laboratory for Molecular Sciences, State Key Laboratory for Structural Chemistry of Unstable and Stable Species, CAS Research/Education Center for Excellence in Molecular Sciences, Institute of Chemistry, Chinese Academy of Sciences, Zhongguancun, Beijing 100190, China, and University of Chinese Academy of Sciences, Beijing 100049, China}

\author{Victor S. Batista}\email{victor.batista@yale.edu}
\affiliation{Department of Chemistry, Yale University, New Haven, CT 06520, USA}
\affiliation{Yale Quantum Institute, Yale University, New Haven, CT 06511, USA}

\begin{abstract}
Atomic and molecular scattering at semiconductor interfaces plays a central role in surface chemistry and catalysis, yet predictive simulations remain challenging due to strong nonadiabatic effects causing the breakdown of the Born–Oppenheimer approximation. Here, we present fully quantum simulations of H-atom scattering from the Ge(111)c(2$\times$8) rest site using the hierarchical equations of motion (HEOM) with matrix product states (MPS). The system is modeled by mapping a density functional theory (DFT) potential energy surface onto a Newns--Anderson Hamiltonian. 
Our simulations reproduce the experimentally observed bimodal kinetic energy distributions, capturing both elastic and energy-loss channels. By systematically examining atom--surface coupling, incident energy, and isotope substitution, we identify the strong-coupling regime required to recover the experimental energy-loss profile. This regime suppresses the elastic peak, implying additional site-specific scattering channels in the observed elastic peak. Deuterium substitution further produces a subtle shift in the energy-loss peak, consistent with experiment. 
These results establish HEOM as a rigorous framework for quantum surface scattering, capable of capturing nonadiabatic dynamics beyond electronic friction and perturbative approaches. 
\end{abstract}
\maketitle

\section{Introduction}

The Born--Oppenheimer approximation (BOA)\cite{born27} is a cornerstone of modern quantum chemistry. It assumes a separation of nuclear and electronic motion, with nuclei evolving on a potential energy surface determined by the electrons. The BOA has been so successful that it is often regarded as a paradigm of molecular science. Despite its success, the BOA breaks down in many important scenarios, including nonadiabatic transitions at conical intersections,\cite{domcke04,schuurman18} polaron formation in solids,\cite{li13b,chen15a,franchini21} and ultrafast photoinduced dynamics in chemical and biological systems.\cite{sun18,tong20,engel07,ishizaki09b,scholes17,dan25} Another critical case arises in molecular interactions with solid surfaces, where the dense manifold of surface electronic states necessitates explicit treatment of electronic excitations in nuclear dynamics. Since such processes are central to heterogeneous catalysis, understanding them is essential for advancing chemical theory and optimizing catalytic processes.\cite{wodtke16}  

Scattering experiments provide a direct quantification of the  interactions of atoms and molecules with solid surfaces. The simplest atom, H atom, has already revealed numerous nonadiabatic effects beyond the BOA in scattering experiments with metal surfaces.
\cite{kandratsenka18,bunermann15,dorenkamp18,alducin17} 
While momentum and energy conservation predict that in a binary elastic collision with a much heavier atom an H atom would rebound with nearly its initial velocity--—valid for insulating surfaces--—scattering from a clean Au(111) surface shows pronounced translational energy loss.\cite{bunermann15} This loss originates from electron transfer and the creation of electron--hole pairs in the metal, which carry away energy from the scattered atom.  

These observations of scattering experiments can be explained by the electronic friction theory, which approximates the effect of surface excitations as frictional and stochastic forces acting on nuclei.\cite{head-gordon95,krishna06,dou18,maurer16,wodtke16,dorenkamp18,spiering18,martin22}
Within this framework, nuclei evolve on the adiabatic BOA surface, while nonadiabatic effects enter perturbatively. The electronic friction theory provides an explicit expression for the friction coefficient, which can be computed using \textit{ab initio} methods. 
For H-atom scattering from Au(111), the local density friction approximation (LDFA)\cite{li92} combined with Langevin molecular dynamics successfully reproduced experimental energy-loss distributions.\cite{bunermann15}  

However, electronic friction has been found to be valid only when nonadiabatic effects primarily involve low-energy excitations. In this perturbative regime, deviations from BOA are modest. When high-energy excitations dominate, the theory fails to make reliable predictions.\cite{kandratsenka18,box21,gardner23,kruger23,malpathak24} This failure is more pronounced on semiconductors, where electron--hole pair excitations require energies above the band gap. In particular, Kr\"uger et al.\cite{kruger23,kruger24} observed bimodal energy-loss distributions in H-atom scattering from Ge(111)c(2$\times$8), revealing two distinct channels. One channel reflected adiabatic dynamics well described by BOA molecular dynamics, while the second exhibited an energy-loss onset at the Ge band gap. In contrast, the electronic friction theory predicted only one single peak between the two experimental peaks, and failed to capture the correct onset of energy loss at the gap energy.

Recent theoretical studies have provided further insights. Zhu et al.\cite{zhu24} combined density functional theory (DFT) with time-dependent nonadiabatic molecular dynamics (NAMD), showing that nonadiabatic
electronic transitions are site-specific and occur selectively at rest sites within a specific spin manifold, although the single-electron treatment limits quantitative comparison with experiment. Lu et al.\cite{lu25} examined independent electron surface hopping (IESH) and Ehrenfest dynamics, finding that Ehrenfest incorrectly predicts energy loss even for incident energies below the band gap, whereas IESH correctly enforces the gap threshold, with the minimum energy loss equal to the band gap, but the resulting energy-loss profiles increasingly deviate from experiment at higher incident energies.

Here, we employ the hierarchical equations of motion (HEOM) method\cite{tanimura89,tanimura06,jin07,jin08} to investigate the scattering dynamics of the H atom on Ge(111)c(2$\times$8) surfaces. HEOM is a non-perturbative, numerically exact framework for open quantum systems that has been widely applied to molecular--metal problems.\cite{hartle15b,schinabeck16,erpenbeck19,erpenbeck20,zhang20,ke21,kaspar22,preston25} Unlike approximate approaches such as electronic friction theory, HEOM rigorously accounts for atom--surface interactions and captures strong nonadiabatic effects. Its high computational cost, however, has historically limited its application to realistic systems. 

Two recent advances have enabled the application of HEOM to simulate scattering dynamics on semiconductor surfaces.
First, with the remarkable success of tensor network
techniques in representing high-dimensional quantum states,
matrix product state (MPS) representations have been
introduced into HEOM,\cite{shi18,ke22,guan24,zhang25} greatly
improving its computational efficiency and enabling
applications to more complex scenarios. In particular, the
recently developed \texttt{mpsqd} package by Guan et
al.\cite{guan24} provides an efficient and user-friendly
implementation of MPS-HEOM, substantially facilitating
practical simulations. 
More recently, Preston et al.\cite{preston25} used MPS-HEOM to investigate vibrational energy relaxation in NO scattering from 
the Au(111) surface, another system where strong nonadiabatic effects render electronic friction theory inadequate. Zhang et al.\cite{zhang25} further developed a multiset MPS-HEOM approach to provide a more detailed study on the effect of molecule-metal coupling strength, and compared with the scattering of NO from the Ag(111) surface.

Second, HEOM decomposes environmental interactions into a
finite set of effective modes, and the computational cost
scales exponentially with the number of effective modes.
Traditionally, this has limited HEOM to relatively simple
environmental structures. For semiconductors, the complex
band structures pose additional challenges for such
decompositions. Significant efforts have been made to reduce
the number of effective modes. Recently, our group
introduced the barycentric spectrum decomposition (BSD)
method,\cite{xu22,dan23} which employs the adaptive
Antoulas–Anderson (AAA) algorithm
\cite{nakatsukasa18,hofreither21} based on barycentric
rational representation to the decomposition of effective
modes. BSD has been shown to handle fermionic reservoirs
effectively and is applicable to, in principle, arbitrary
band structures.\cite{dan23,takahashi24}

In this work, we adopt a Newns--Anderson Hamiltonian to describe H-atom scattering on Ge(111), treating both nuclear motion and surface electronic degrees of freedom quantum mechanically. Motivated by experimental and theoretical evidence,\cite{kruger24,zhu24} we focus on the Ge(111) rest site, where nonadiabatic electronic transitions are most likely to occur. This choice captures the dominant scattering pathway while keeping the HEOM simulations tractable.

The one-dimensional potential energy surface (PES) for the Ge(111) rest site is obtained from DFT calculations and fitted to the adiabatic PES of the Newns--Anderson model to determine the model parameters. The resulting Hamiltonian is solved using the MPS-HEOM approach, enabling a fully quantum investigation of how the scattering dynamics depend on atom–-surface coupling strength, incident energy, and isotope substitution.

Our simulations reveal that realistic scattering occurs in the strong-coupling regime, where the energy-loss channel closely matches experimental observations. The results further support the interpretation that the two observed scattering channels originate from collisions at different surface sites, consistent with the experimentally proposed site-specific mechanism. In addition, isotope substitution produces a systematic shift of the deuterium energy-loss peak toward higher energy, consistent with the experimental result.

The remainder of this paper is organized as follows. Section~\ref{sec:theory} outlines the theoretical framework, including the Newns--Anderson model and the HEOM methodology. Section~\ref{sec:comp_setup} describes the computational setup, where the adiabatic PES is fitted to DFT results to determine model parameters and define the initial kinetic energy distributions. Section~\ref{sec:result} presents the scattering dynamics and examines the effects of atom–surface coupling strength, incident energy, and isotope substitution. Section~\ref{sec:Discussion} compares our simulations with previous experimental observations and provides additional physical insight. Finally, Section~\ref{sec:Conclusion} summarizes the conclusions of this work.

\section{Theory}\label{sec:theory}
\subsection{Newns--Anderson Model}

As in many previous studies,\cite{anderson61,newns69,schmickler86,
head-gordon95,schmickler02,nitzan01,dou17} we employ a Newns--Anderson type Hamiltonian to describe the interaction of a hydrogen atom (the ``system") with the continuum of electronic states of the Ge(111)c(2$\times$8) surface (the ``bath"). The total Hamiltonian is given by (with $\hbar = 1$ throughout):
\begin{equation}
\label{Eq:hamtot}
H = \frac{{\bf p}^2}{2M} 
+ U_0({\bf x}) \hat{c}_H^\dagger \hat{c}_H
+ U_a({\bf x}) \left(1-\hat{c}_H^\dagger \hat{c}_H\right)
+ \sum_n \epsilon_n \hat{c}_n^\dagger \hat{c}_n
+ \sum_n g({\bf x}) \left(V_n \hat{c}_n^\dagger \hat{c}_H 
+ V_n^* \hat{c}_H^\dagger \hat{c}_n \right) \; ,
\end{equation}
where $\hat{c}_H^\dagger$ and $\hat{c}_H$ are the creation and annihilation operators of the H atom electronic state, while $\hat{c}_n^\dagger$ and $\hat{c}_n$ correspond to the surface electronic state $\ket{n}$ with energy $\epsilon_n$. The H atom nuclear degrees of freedom are described by momentum ${\bf p}$, coordinate ${\bf x}$, and mass $M$.  

$U_0({\bf x})$ and $U_a({\bf x})$ denote the diabatic potential energy surfaces (PESs) governing nuclear motion of the H atom. $U_0({\bf x})$ corresponds to the ground state ($\ket{0}$ state) of the neutral H atom, whereas $U_a({\bf x})$ represents the state after electron transfer between the H atom and the Ge surface ($\ket{a}$ state). We choose $\ket{a}$ as the H cation state (so that $\hat{c}_H^\dagger \ket{a} = \ket{0}$), consistent with reports that when H atoms adsorb on the
rest atom of the Ge surface, the H atom loses its
electron.\cite{klitsner91,razado06,zhu24} 
Nevertheless, while we focus here on the H$^+$ mechanism at the rest site, all conclusions of this work equally apply to the H$^-$ mechanism. This is due to the inherent symmetry under the interchange of $\hat{c}_H$ and $\hat{c}_H^\dagger$ in the Newns–Anderson model, which makes the dynamics the same whether $\ket{a}$ is assigned as H$^+$ or H$^-$.
The function $g({\bf x})$ accounts for the position-dependent H--surface interaction, and $V_n$ denotes the coupling constant between the surface continuum electronic state $\ket{n}$ and the H atom electronic state.  

The system--bath coupling is characterized by the hybridization function,  
\begin{equation}
\Gamma(\epsilon) = 2\pi \sum_{n} |V_n|^{2} \, \delta(\epsilon-\epsilon_n) \; ,
\end{equation}
which encodes the interaction between the H atom and the Ge surface electronic continuum.  

\subsection{Adiabatic Potential Energy Surface}
\label{sec:Eadia} 

To obtain a physically meaningful parameterization of the Newns--Anderson model, we fit its adiabatic potential energy surface (PES) to density functional theory (DFT) results. In our previous work,\cite{dan23b} we developed a method to compute the adiabatic potential energy surfaces of the Newns--Anderson model based on the Hellmann--Feynman theorem. The key steps are summarized below.  

First, fixing the nuclear position ${\bf x}$, the electronic Hamiltonian of Eq.~(\ref{Eq:hamtot}) reduces to  
\begin{equation}\label{Eq:ele_hamt}
H_e({\bf x}) = U_a({\bf x}) + \big[ U_0({\bf x}) - U_a({\bf x}) \big] 
\hat{c}_H^\dagger \hat{c}_H 
+ \sum_n \epsilon_n \hat{c}_n^\dagger \hat{c}_n 
+ \sum_n \Big( V_n({\bf x}) \hat{c}_n^\dagger \hat{c}_H 
+ V_n^*({\bf x}) \hat{c}_H^\dagger \hat{c}_n \Big) \; ,
\end{equation}
where $V_n({\bf x}) = g({\bf x}) V_n$. This corresponds to a resonant-level model (RLM) with a hybridization function  
\begin{equation}
\Gamma(\epsilon,{\bf x}) = g^2({\bf x}) \, \Gamma(\epsilon) \; ,
\end{equation}
which can be solved exactly using the Green’s function formalism.\cite{mahan00,haug08}  

The retarded Green’s function is defined as\cite{haug08,dan22}  
\begin{equation}\label{Eq:green_RLM}
\begin{split}
G^r(\epsilon,{\bf x}) &= \Big[\epsilon - \epsilon_H - \Sigma^r(\epsilon,{\bf x}) \Big]^{-1}, \\
\Sigma^r(\epsilon,{\bf x}) &= \int \frac{d\epsilon'}{2\pi} 
\frac{\Gamma(\epsilon',{\bf x})}{\epsilon - \epsilon' + i0^+} \; ,
\end{split}
\end{equation}
where $\epsilon_H = U_0({\bf x}) - U_a({\bf x})$ and ${\Sigma}^r(\epsilon,{\bf x})$ is the electronic self-energy.  

With $G^r(\epsilon,{\bf x})$ obtained, the H-state occupation  
\begin{equation}
\langle n_H \rangle({\bf x}) = \langle \hat{c}_H^\dagger \hat{c}_H \rangle
\end{equation}
and the interaction energy  
\begin{equation}
\langle H_{\text{int}} \rangle({\bf x}) 
= \Big\langle \sum_n \big( V_n({\bf x}) \hat{c}_n^\dagger \hat{c}_H 
+ V_n^*({\bf x}) \hat{c}_H^\dagger \hat{c}_n \big) \Big\rangle
\end{equation}
can be expressed as\cite{schmickler86,muscat78,mahan00,haug08,dan23b}  
\begin{equation}\label{Eq:occu_int}
\begin{split}
\langle n_H \rangle({\bf x}) &= -\frac{1}{\pi} 
\int_{-\infty}^{\infty} d\epsilon \,
f(\epsilon)\, \mathrm{Im}[G^r(\epsilon,{\bf x})], \\
\langle H_{\text{int}} \rangle({\bf x}) &= -\frac{2}{\pi} 
\int_{-\infty}^{\infty} d\epsilon \,
f(\epsilon)\, \mathrm{Im}[G^r(\epsilon,{\bf x}) \, \Sigma^r(\epsilon,{\bf x})] \; ,
\end{split}
\end{equation}
where $\mathrm{Im}[\cdot]$ denotes the imaginary part and  
$f(\epsilon) = [1 + e^{\beta(\epsilon - \mu)}]^{-1}$ is the Fermi–Dirac distribution with $\beta = 1/k_B T$ and chemical potential $\mu$.  

The free energy of $H_e({\bf x})$ in the grand canonical ensemble is defined as  
\begin{equation}
\mathcal{G}({\bf x}) = -k_B T \ln \Xi({\bf x}), 
\qquad 
\Xi({\bf x}) = \mathrm{Tr}\, e^{-\beta(H_e({\bf x}) - \mu \hat{N})}.
\end{equation}
Using the procedure in Ref.~\citenum{dan23b}, the free energy can be expressed as  
\begin{equation}\label{Eq:free_ene}
\mathcal{G}({\bf x})
= U_a({\bf x}) - U_a({\bf x}_{\mathrm{ref}})
+ \int_{{\bf x}_{\mathrm{ref}}}^{\bf x} d{\bf x}' 
\Bigg\{ 
\frac{\partial [U_0({\bf x}') - U_a({\bf x}')]}{\partial {\bf x}'} 
\, \langle n_H \rangle({\bf x}')
+ \frac{1}{g({\bf x}')} \frac{\partial g({\bf x}')}{\partial {\bf x}'} 
\, \langle H_{\text{int}} \rangle({\bf x}')
\Bigg\},
\end{equation}
where ${\bf x}_{\mathrm{ref}}$ is a reference position far from the surface.  

At zero temperature, the free energy $\mathcal{G}({\bf x})$ reduces to the adiabatic potential energy surface of $H_e({\bf x})$. We therefore denote  
\begin{equation}\label{Eq:adia_ene}
E_{\text{adia}}({\bf x}) \equiv \mathcal{G}({\bf x}) \big|_{T=0} \; ,
\end{equation}
as the adiabatic PES of the Newns--Anderson model.

\subsection{HEOM in the MPS Framework}
\label{sec:heom_mps}

The Newns--Anderson Hamiltonian in Eq.~(\ref{Eq:hamtot}) describes an open quantum system, where the H atom (system) interacts with a fermionic environment consisting of the continuum of electronic states in the semiconductor surface. The dynamics can, in principle, be solved exactly using the HEOM method.\cite{jin08,hartle13,xu19}  

For semiconductors, however, the complex band structure makes traditional HEOM implementations prohibitively expensive. Shi \textit{et al.}\cite{shi18} demonstrated that tensor network methods, in particular matrix product states (MPS), provide a highly efficient representation of HEOM, significantly reducing computational cost and extending its applicability. Here, we describe the bath decomposition, HEOM formalism, and MPS representation for MPS-HEOM propagation.

\subsubsection{Bath Decomposition}  
Within HEOM, the effect of the electronic continuum is encoded in a set of effective modes obtained by decomposing the bath correlation function into a sum of exponentials:  
\begin{equation}\label{Eq:SOP_corr}
C^\sigma(t) = \int_{-\infty}^{+\infty} \frac{d\epsilon}{2\pi} 
\, e^{\sigma i\epsilon t}\,\Gamma(\epsilon)\, f^\sigma(\epsilon) 
\simeq \sum_{k=1}^K d^\sigma_{k} \, e^{-\nu^\sigma_{k}t} \; ,
\end{equation}
where $\sigma \in \{+,-\}$ distinguishes electrons $(+)$ and holes $(-)$, $f^\sigma(\epsilon) = [1 + e^{\sigma \beta(\epsilon-\mu)}]^{-1}$ is the Fermi distribution, and $\Gamma(\epsilon)$ is the hybridization function. The coefficients $d^\sigma_k$ and frequencies $\nu^\sigma_k$ define the effective modes. The decomposition of $C^\sigma(t)$ for semiconductor-like $\Gamma(\epsilon)$ is performed using the recently developed BSD method.\cite{dan23} 

\subsubsection{HEOM Formalism}  
After decomposition, the system coupled to an infinite bath is reformulated as a system interacting with a finite set of effective modes. The density operator $\hat{\rho}_{\bf J}$ is defined in the composite space  
\begin{equation}\label{Eq:heom_space}
    \hat{\rho}_{\bf{J}} \in \mathcal{F}_e \otimes \mathcal{F}_e^* 
    \otimes \mathcal{H}_x \otimes \mathcal{H}_x^* 
    \otimes \mathcal{F}_{\rm eff} ,
\end{equation}
where $\mathcal{F}_e$ and $\mathcal{F}_e^*$ are the electronic Fock space ($\ket{0}$ and $\ket{a}$) of the H atom and its dual, $\mathcal{H}_x$ and $\mathcal{H}_x^*$ are the nuclear coordinate space of the H atom and its dual, and $\mathcal{F}_{\rm eff}=\{ {\bf J}
\}$ contains all possible effective mode configurations, 
where each ${\bf J}=\{j_{1}j_{2},...,j_{\Omega}\}$ represents a specific
configuration of occupied effective modes with a given
order. Here, each $j=(k,\sigma)$, and the maximum number of
occupied modes $\Omega$ is $2K$.  

The reduced density matrix of the system is $\hat{\rho}_{\bf 0}$, corresponding to the configuration where all effective modes are unoccupied. For the Newns--Anderson model with a position-dependent atom--surface coupling in Eq.~(\ref{Eq:hamtot}), the HEOM reads:\cite{erpenbeck18b,ke21,dan23b}  
\begin{equation}
\label{Eq:drho1}
\frac{\partial}{\partial t}\hat{\rho}_{\bf J}(t)
= -i[H_{\rm mol},\hat{\rho}_{\bf J}]
-\gamma_{\bf{J}} \hat{\rho}_{\bf J}
-i\sum_{m=1}^{\Omega} (-1)^{\Omega-m} 
\mathcal{C}_{j_m}\hat{\rho}_{{\bf J}_m^-}
- i\sum_{j_{\Omega+1}} 
\mathcal{A}_{\bar{j}_{\Omega+1}} \hat{\rho}_{{\bf J}^+} \,,
\end{equation}
where $\gamma_{\bf{J}} = \sum\limits_{( k, \sigma)
\in {\bf{J}}}\nu^\sigma_{k} $.
${{\bf J}^{+}}$ denotes $\{j_{1}j_{2},...,j_{\Omega},j_{\Omega+1}\}$, 
and ${{{\bf J}_{m}^{-}}}$ denotes 
$\{j_{1},...,j_{m-1},j_{m+1},...,j_{\Omega}\}$, 
$\bar{j}=( k, \bar{\sigma})$ with $\bar{\sigma}=-\sigma$. 
The molecular Hamiltonian is the system part of Eq.~(\ref{Eq:hamtot}):  
\begin{equation}
H_{\rm mol} = \frac{{\bf p}^2}{2M} + U_0({\bf x}) \hat{c}_H^\dagger \hat{c}_H 
+ U_a({\bf x}) \left(1-\hat{c}_H^\dagger \hat{c}_H\right).
\end{equation}  

The operators $\mathcal{C}_j$ and $\mathcal{A}_j$ act as
“creation” and “annihilation” superoperators for the
effective modes, coupling $\hat{\rho}_{\bf J}$ to
$\hat{\rho}_{{\bf J}_m^-}$ and $\hat{\rho}_{{\bf J}^+}$,
respectively. Their explicit actions are given by: 
\begin{subequations}\label{Eq:ACoperator}
\begin{equation}
\mathcal{C}_j\hat{\rho}_{\bf J} = 
g({\bf x}) d_{j}\hat{c}_j \hat{\rho}_{\bf J} - 
(-1)^{\Omega} d^*_{\bar{j}} \hat{\rho}_{\bf J} \hat{c}_j g({\bf x}),
\end{equation}
\begin{equation}
\mathcal{A}_j\hat{\rho}_{\bf J} = 
g({\bf x}) \hat{c}_j \hat{\rho}_{\bf J} + 
(-1)^{\Omega} \hat{\rho}_{\bf J} \hat{c}_j g({\bf x}) .
\end{equation}
\end{subequations}
For $j = ( k, \sigma)$, we have $d_{j}=d^\sigma_{k}$, and
$\hat c_j = \hat c^\sigma_H$ where $\hat c^{-(+)}_H \equiv
\hat c^{(\dag)}_H$ denotes the system annihilation
(creation) operators.

\subsubsection{MPS Representation}  
In practice, the large hierarchy of all density operators $\hat{\rho}_{\bf J}$ is represented compactly using the MPS ansatz:\cite{shi18,guan24}  
\begin{equation}
    \hat{\rho}_{\bf J} \approx 
    \sum_{\alpha_1=1}^{r_1} \cdots \sum_{\alpha_{d-1}=1}^{r_{d-1}}
    A_1(n_1,\alpha_1) A_2(\alpha_1,n_2,\alpha_2) \cdots 
    A_d(\alpha_{d-1},n_d),
\end{equation}
here, $A_i$ are three-dimensional tensors, except that the first and last tensors are two-dimensional. $r_i \, (i=1,...,d)$ denotes the $i$-th bond dimension of the MPS. The total dimension is $d=2K+4$, covering all
spaces included in Eq.~(\ref{Eq:heom_space}). 
Specifically,
the first four tensors correspond to the degrees of freedom
for the electronic Fock space, its dual, the nuclear
coordinate space of the H atom, and its dual,
respectively. The remaining $2K$ tensors correspond to the
degrees of freedom in $\mathcal{F}_{\rm eff}$. In detail, $n_1$
and $n_2$ represent the electronic states $\{0,a\}$, while
$n_3$ and $n_4$ correspond to the nuclear coordinate basis
$\{\bf x\}$. All possible configurations $\bf J$ in
$\mathcal{F}_{\rm eff}$ are mapped to the occupation number
basis $\{n_5,\cdots,n_{2K+4}\}$, where $n_i=0$ or 1 for
$i=5,...,2K+4$.

Time evolution is performed with the time-dependent variational principle (TDVP),\cite{lubich15} which integrates the dynamics while maintaining a fixed MPS bond dimension. 
In practice, simulations are performed by systematically increasing the bond dimension to ensure convergence of the dynamics. Further details can be found in Refs.~\citenum{shi18} and \citenum{guan24}. The TDVP-based
HEOM propagation has been implemented in the \texttt{mpsqd}\cite{guan24} package.

\section{Computational Setup}\label{sec:comp_setup}

\subsection{System--Bath Interaction}

In this work, the coupling between the H atom and the Ge(111) surface is described by the following hybridization function:  
\begin{equation}\label{Eq:level_width_func}
\Gamma(\epsilon) = 
\frac{\eta \gamma^2}{(\epsilon-\epsilon_0)^2 + \gamma^2}
\left[ 1 - \frac{1}{1+e^{(\epsilon-E_B/2)/\delta}}
          \frac{1}{1+e^{-(\epsilon+E_B/2)/\delta}} \right],
\end{equation}
where $E_B = 0.49$~eV is the Ge(111) band gap,\cite{kruger23} $\delta = 0.02$~eV controls the smoothness of the band edges, $\epsilon_0 = 0$ sets the band center, and $\gamma = 1.5$~eV determines the bandwidth, 
which is chosen to mimic the two peaks of the surface density of states near the band edges.\cite{feenstra06} Although this choice may not exactly match the real system, tests presented in the Supporting Information demonstrate that the results reported here are insensitive to the specific value of the bandwidth.
The overall coupling strength $\eta$ is set to 3 eV, and the rationale for this choice will be discussed in Sec.~\ref{sec:Dis_sim_vs_exp}.

Physically, $\Gamma(\epsilon)$ characterizes the energy-dependent interaction strength between the H atom electronic state and the continuum of surface states, thereby determining the probability of charge transfer and energy dissipation into the surface. This hybridization function, together with the Fermi distribution, serves as input for the bath correlation function $C^\sigma(t)$ [Eq.~(\ref{Eq:SOP_corr})]. The temperature is set to $T=300$~K, which enters via the Fermi function. To decompose $C^\sigma(t)$ into exponential terms, we employ the BSD scheme,\cite{dan23} which expands both the hybridization function and the Fermi distribution into a finite set of effective modes.  

Figure~\ref{fig:bath_bsd} illustrates the BSD approximations for $\Gamma(\epsilon)$ and the Fermi distribution. The relative error of their product $\Gamma(\epsilon)f^\sigma(\epsilon)$ remains below $10^{-3}$ across the entire energy range. Such accuracy has been demonstrated to be sufficient for the present type of simulations.\cite{dan23} In practice, 13 effective modes are obtained: 6 from the hybridization function and 7 from the Fermi distribution.  

\begin{figure}[h!]
\centering
\includegraphics[width=0.8\linewidth]{Fig_bath_bsd.eps}
\caption{Accuracy of the BSD approximation for (a) the hybridization
function $\Gamma(\epsilon)$ (normalized as $\Gamma(\epsilon)/\eta$) and (b) the
Fermi distribution at $T=300$~K. The relative error of their product
$\Gamma(\epsilon)f^\sigma(\epsilon)$ is below $10^{-3}$ across the full energy range.}
\label{fig:bath_bsd}
\end{figure}

\subsection{Model Parametrization}

The potential energy surface (PES) for H-atom scattering at the Ge(111) rest site was obtained from density functional theory (DFT) calculations using the Vienna \textit{Ab Initio} Simulation Package (VASP) with the projector augmented wave (PAW) method.\cite{blochl94} Calculations were carried out at the PBE level\cite{perdew96} with D3-BJ dispersion corrections.\cite{grimme10,grimme11} A plane-wave cutoff of 450 eV and a $\Gamma$-centered $1 \times 3 \times 1$ Monkhorst--Pack $k$-point grid\cite{monkhorst76} were employed. The Ge(111) slab~\cite{chadi81} was modeled with four atomic layers, with the bottom two layers fixed during geometry optimization.  

To reproduce the DFT PES within the Newns--Anderson model, we fit the adiabatic PES using the following analytic forms for $U_0(x)$ and $U_a(x)$:  
\begin{subequations}\label{Eq:syspot}
\begin{equation}
U_0(x) = A_0 \left(e^{2C_0(x-x_0)} - 2e^{C_0(x-x_0)}\right),
\end{equation}
\begin{equation}
U_a(x) = A_1 \left(e^{2C_1(x-x_1)} - 2e^{C_1(x-x_1)}\right) + B_1 + \frac{1}{(x-D_1)^4}.
\end{equation}
\end{subequations}
Here, $U_0(x)$ follows the Morse potential form, while $U_a(x)$ includes an additional short-range repulsive term $1/(x-D_1)^4$ to capture the strong repulsion experienced by the H atom near the Ge(111) surface. The constant offset $B_1$ is fixed at 8.8 eV, corresponding to the difference between the H atom ionization potential (13.6 eV) and the Ge(111) work function (4.8 eV).\cite{gobeli64}  

The position-dependent interaction between the H atom and the surface is described by a sigmoidal coupling function,  
\begin{equation}
g(x) = \frac{2}{1+e^{-c_g(x-x_g)}},
\end{equation}
with $x_g$ and $c_g$ as adjustable parameters.  

Using the procedure in Sec.~\ref{sec:Eadia}, we calculate the adiabatic energy $E_{\rm adia}(x)$ and occupation number $\langle n_H\rangle(x)$ for a given parameter set. These quantities are then fitted to DFT reference data: the DFT PES and the spin magnetic moment. The resulting optimized parameters are summarized in Table~\ref{Tab:para_all}.  

Figure~\ref{fig:allpes}(a) shows the model potentials $U_0(x)$, $U_a(x)$, and the coupling function $g(x)$ along with the computed $E_{\rm adia}(x)$ and the DFT PES. Figure~\ref{fig:allpes}(b) compares the calculated occupation number $\langle n_H\rangle(x)$ with the spin magnetic moment obtained from DFT. 
\begin{figure}[h!]
\centering
\includegraphics[width=0.6\linewidth]{Fig_allPES.eps}
\caption{(a) Model potentials $U_0(x)$, $U_a(x)$, and position-dependent coupling $g(x)$ in the Newns--Anderson model, together with the computed adiabatic energy $E_{\rm adia}(x)$ and DFT reference $E_{\rm DFT}$. (b) Calculated occupation number $\langle n_H\rangle(x)$ compared with the spin magnetic moment $\mu_{H}$ from DFT results. All parameters are in Table~\ref{Tab:para_all}.
}
\label{fig:allpes}
\end{figure}

Overall, the Newns--Anderson model captures the essential features of the DFT PES and the spin magnetic moment transition. While minor deviations arise from the use of analytical functional forms, our aim is not to reproduce the DFT PES exactly. Rather, the resulting parameter set provides a physically meaningful basis for simulating the scattering dynamics.

\subsection{Kinetic Energy Distribution}\label{sec:proj}

The momentum-space distribution at electronic state $\ket{i}$ ($i \in \{0,a\}$) is defined as  
\begin{equation}\label{Eq:k-dis}
\rho_{ii}(k) = \frac{1}{2\pi} \int dx \int dx' \, e^{-ik(x-x')} \, \rho_{ii}(x,x') ,
\end{equation}
where $\rho_{ii}(x,x')$ is the reduced coordinate-space density matrix at state $i$:  
\begin{equation}
\rho_{ii}(x,x') \equiv \bra{i}\bra{x} \hat{\rho} \ket{x'}\ket{i}.
\end{equation}
Here, $\hat{\rho}$ is the reduced density operator (i.e., $\hat{\rho}_{\bf 0}$ in HEOM). In other words, $\rho_{ii}(k)$ corresponds to the diagonal element of the Fourier transform of the coordinate-space reduced density operator. 

The kinetic energy distribution at state $\ket{i}$, denoted as 
$P_i(E)$, is then given by  
\begin{equation}\label{Eq:kinetic_energy_dis}
P_i(E) = \rho_{ii}(k)\frac{dk}{dE} 
= \frac{\rho_{ii}(k)}{\hbar} \sqrt{\frac{M}{2E}} ,
\end{equation}
where we use the relation $E = \hbar^2 k^2 / (2M)$ for the kinetic energy.

As $E \to 0$, $\sqrt{1/E}$ becomes large, resulting in a
sharp peak in the numerically obtained $P_i(E)$ at small $E$
values.\cite{dan23b} According to our PES shown in
Fig.~\ref{fig:allpes}, H atoms with such low kinetic
energies cannot escape from the semiconductor surface and
remain trapped nearby. In contrast, experimental kinetic
energy distributions, measured using time-of-flight
techniques,\cite{kruger23} should not include contributions
from atoms trapped near the surface.

To enable a direct comparison with experiments, we exclude
these trapped components when calculating the kinetic energy
distribution. This is achieved by projecting out the
near-surface contributions: 
% Based on this consideration, we project out the near-surface contributions when calculating the kinetic energy distribution by defining
\begin{equation}
\hat{\rho}^{\mathrm{pj}} = \mathcal{P}\hat{\rho}  \mathcal{P} \;\;,
\end{equation}
where $\hat{\rho}$ is the reduced density operator and
$\mathcal{P}$ is a projection operator that removes the
distribution near the surface. In this work, we choose the
projection operator as \begin{equation} \mathcal{P} = I -
\sum_{E_i<0} |\psi_i\rangle\langle\psi_i|
\end{equation} 
where $I$ is the identity operator, $|\psi_i\rangle$ and
$E_i$ are the $i$-th eigenstate and eigenenergy of the
adiabatic PES, i.e., $E_{\rm adia}(x)$ in
Eq.~(\ref{Eq:adia_ene}). 
By applying this projection operator, contributions with $E_i<0$ are eliminated, effectively removing the trapped components that do not have sufficient energy to escape the trapping region.
 
By performing the same procedure as in
Eqs.~(\ref{Eq:k-dis})–(\ref{Eq:kinetic_energy_dis}) but
using $\hat{\rho}^{\mathrm{pj}}$, we obtain the kinetic
energy distribution with the bound state contributions
projected out, denoted as $P_i^{\mathrm{pj}} (E)$.

\section{Results}\label{sec:result}

\subsection{H-Atom Scattering Dynamics}

In all simulations, the nuclear degree of freedom $x$ is represented using the potential optimized discrete variable representation (PO-DVR).\cite{echave92,colbert92} Unless otherwise noted, the DVR grid spans $x \in [-60,\,0.2]$~a.u. The number of PO-DVR basis functions, up to 1000, depends on the incident kinetic energy, atom--surface coupling strength, and isotope effect. 
The H atom is initially prepared in the neutral state $\ket{0}$ with a Gaussian wavepacket:
\begin{equation}
    \Psi^{\rm ini}(x) = \frac{1}{(2\pi\sigma_x^2)^{\frac{1}{4}}} 
\exp\Big[ -\frac{(x-x_{\rm ini})^2}{4\sigma_x^2} + ip_{\rm ini}(x-x_{\rm ini})
 \Big] \;\;,
\end{equation}
centered at $x=-11$~a.u. and a spatial width of $\sigma_x = 1$~a.u. This spatial width corresponds to a Gaussian momentum distribution centered at $p_{\mathrm{ini}}$, whose full width at half maximum (FWHM) in kinetic energy is approximately 300 meV (see Figure~\ref{fig:PE0}).
% This corresponds to a Gaussian initial momentum distribution centered at $p_{ini}$ with variance $\langle (p-p_{ini})^2\rangle=\hbar^2/(4\sigma_x^2)=1/4$~a.u., yielding an initial kinetic-energy full width at half maximum of approximately 300 meV (see Figure~\ref{fig:PE0}). 
Although the kinetic-energy distributions of atomic hydrogen beams in experiments are typically only a few tens of meV,\cite{bunermann15b} we show in the Supporting Information that this broader initial distribution does not significantly affect the resulting dynamics and does not alter the physical conclusions regarding nonadiabatic scattering drawn in the present work.

The Gaussian wavepacket along the $x$ coordinate implies that, within the HEOM formalism, we initialize the density operators $\hat{\rho}_{\bf J} (t)$ in Eq.~(\ref{Eq:heom_space}) as: 
\begin{equation}
    \hat{\rho}_{\bf J} (t=0) = |0\rangle\langle 0|\otimes |\Psi^{ini}(x)\rangle \langle \Psi^{ini}(x)| \otimes |{\bf 0}\rangle
\end{equation}
where $|{\bf 0}\rangle$ means all effective modes are unoccupied, corresponds to bath thermal equilibrium.

In this subsection, we consider an incident energy $E_{\rm in} = 1.92$~eV. Figure~\ref{fig:elec_pop} shows the population dynamics of the H atom in states $\ket{0}$ and $\ket{a}$ during the scattering process. At $t \approx 20$~fs, the wavepacket reaches the Ge surface and electron transfer from the H atom to the surface begins, seen as a sharp decrease in $P_0(t)$ accompanied by an increase in $P_a(t)$. By $t \approx 30$~fs, the electron transfer reaches its maximum, with nearly the entire wave packet occupying the $|a\rangle$ state. As the atom rebounds, electrons transfer back to the H atom, driving a transition from $|a\rangle$ to $|0\rangle$. Most of this back-transfer is completed by $t \approx 50$ fs, small oscillations in $P_0(t)$ and $P_a(t)$ after 50~fs reflect residual charge exchange near the surface. After $t > 100$~fs, the population curves stabilize. By the end of the simulation, most of the H atom escapes in the $|0\rangle$ state, while a small fraction remains trapped near the surface in the $|a\rangle$ state. 

\begin{figure}[h!]
\centering
\includegraphics[width=0.8\linewidth]{Fig_elec_pop.eps}
\caption{Time evolution of H-atom populations in states $\ket{0}$ and $\ket{a}$, denoted $P_0(t)$ and $P_a(t)$, respectively, during the scattering process.}
\label{fig:elec_pop}
\end{figure}

\begin{figure}[h]
\centering
\includegraphics[width=1\linewidth]{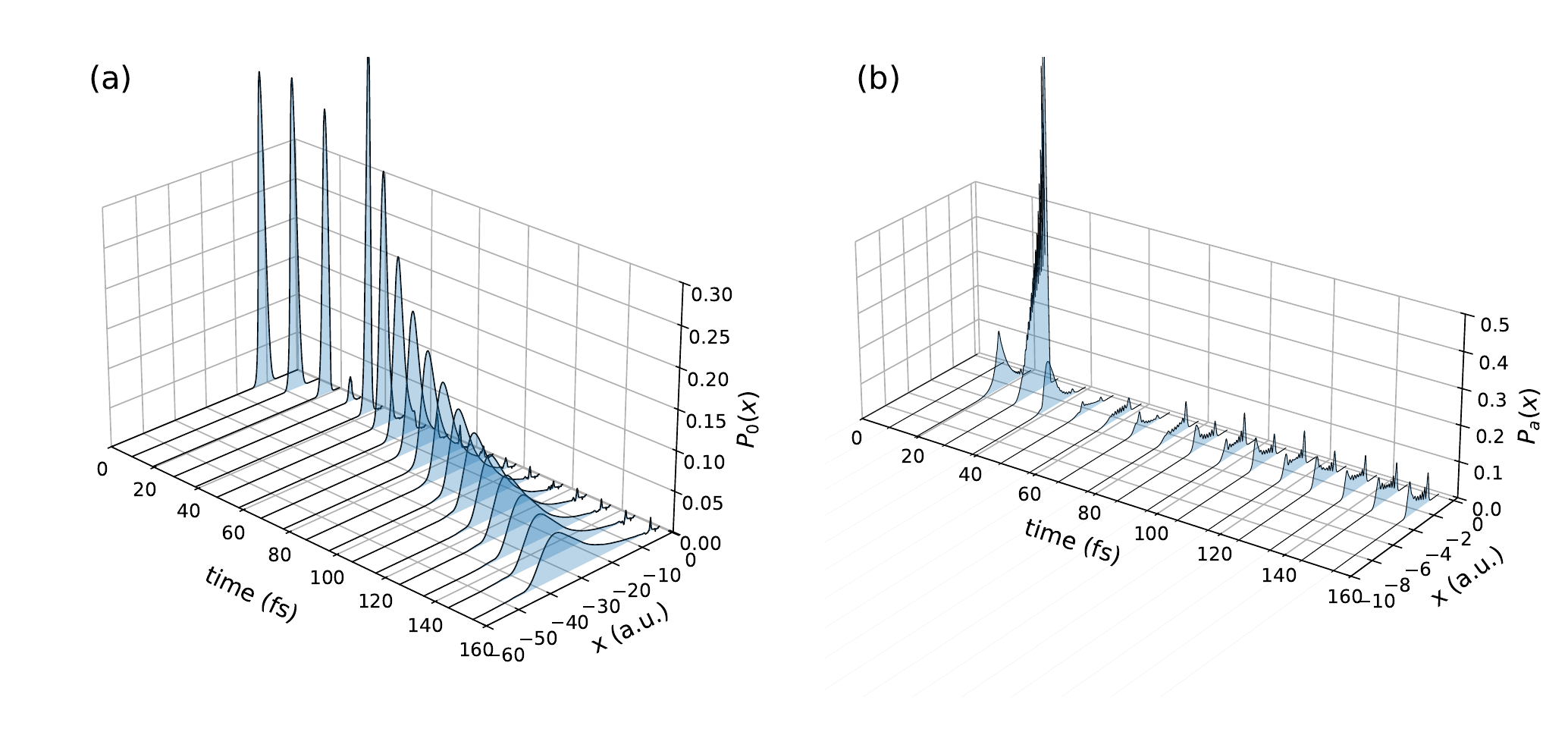}
\caption{Spatial probability distributions of the H atom in the $|0\rangle$ and $|a\rangle$ states during the scattering process, denoted as $P_0(x)$ and $P_a(x)$, respectively. Panel (a) shows $P_0(x)$, and Panel (b) shows $P_a(x)$.}
\label{fig:Pxt_all}
\end{figure}

To further describe the scattering process, we present the evolution of wave packet distributions $P_0(x)$ and $P_a(x)$ in Figure~\ref{fig:Pxt_all}. Initially, the wavepacket approaches the Ge surface from the left ($x < 0$). Upon reaching the surface at $t \approx 20$~fs, electron transfer causes $P_0(x)$ to decrease and $P_a(x)$ to rise. By $t\approx 50$~fs, $P_0(x)$ begins to propagate away from the surface and gradually delocalizes, signaling a loss of kinetic energy.

By the end of the simulation, a small peak of $P_0(x)$ remains localized near the surface, reflecting a low-energy component that lacks sufficient kinetic energy to escape. Meanwhile, a portion of $P_a(x)$ persists, oscillating within the well of $U_a(x)$ near the surface, indicating that some H atoms become adsorbed on the Ge surface in the $|a\rangle$ state after scattering. The pronounced oscillatory features of $P_a(x)$ further suggest that this population occupies excited vibrational levels on the $|a\rangle$ state. 

To facilitate comparison with experimental results, we plot in Figure~\ref{fig:PE0} the kinetic energy distribution of the H atom in the $|0\rangle$ state (H-neutral) at the time of emission ($t=154.8$~fs). 
The projection technique described in Sec.~\ref{sec:proj} is applied to exclude small contributions from wavepackets trapped near the surface. 
The projection removes only a small portion of the low-energy distribution near $E\to 0$, leaving the main peak essentially unaffected. A comparison of the spatial and kinetic energy distributions before and after projection is provided in the Supporting Information.

% The resulting distribution exhibits a bimodal structure. 
% The dominant peak spans 0--1.5~eV with a maximum near 1~eV, corresponding to the energy-loss channel. 
% Its shape and position agree well with the experimental observations reported by Krüger \textit{et al.}\cite{kruger23}
% In addition, a much weaker Gaussian-like peak appears near the incident energy $1.92$~eV. 
% This peak corresponds to the elastic scattering channel, arising from a small portion of the H wavepacket that scatters without exchanging electrons with the surface. 
% Because of its very small weight, this elastic peak is not discernible in Figure~\ref{fig:Pxt_all}. Overall, our simulations capture the bimodal feature observed experimentally. However, the relative ratio of the two channels differ significantly from experiment, which will be discussed later in Sec.~\ref{sec:Discussion}. 

\begin{figure}[h]
\centering
\includegraphics[width=0.8\linewidth]{Fig_PE_t6400.eps}
\caption{Kinetic energy distribution of the H atom emitted in the $|0\rangle$ state at $t=154.8$~fs, projected using the technique in Sec.~\ref{sec:proj} to exclude some near-surface contributions. For comparison, the initial kinetic energy distribution is also plotted (scaled by a factor of 1/3).}
\label{fig:PE0}
\end{figure}

The resulting distribution exhibits a bimodal structure. 
The dominant peak spans 0--1.5~eV with a maximum near 1~eV, 
we refer to this peak as the energy-loss channel.
Because phonon baths are not included in the present Newns–Anderson model, all energy loss arises solely from interactions with the electronic bath.
In addition, a much weaker Gaussian-like peak appears near the incident energy $1.92$~eV. 
It preserves the shape of the initial kinetic-energy distribution and therefore corresponds to essentially no energy loss. We refer to this peak as the energy-loss channel, originating from a small fraction of the H wavepacket that scatters without exchanging electrons with the surface.
Because of its very small weight, this elastic peak is not discernible in Figure~\ref{fig:Pxt_all}.

The dominant energy-loss peak originates from scattering events involving electronic transitions with the semiconductor, thereby reflecting the band structure characteristics of the material.
Consider the following process: an H atom donates an electron to the conduction band of Ge(111), and after the collision, it regains an electron from the valence band. 
Since the returning electron has lower energy than the donated one, such an electronic transition is expected to result in an energy loss approximately equal to the band gap. 
This mechanism explains why the high-energy edge of the loss peak lies $E_g = 0.49$~eV below the incident energy.  

Our elastic scattering channel and energy-loss channel share the same electronic origins as the experimental adiabatic channel and the valence-to-conduction-band (VB–CB) channel reported by Krüger \textit{et al.}\cite{kruger23}, except that phonons are not included in our model.
Overall, the simulations reproduce the experimentally observed bimodal structure, and both the shape and the position of the energy-loss channel agree well with experiment.
However, the relative ratio of the two channels differ substantially.
A comprehensive discussion of these aspects is provided in Sec.~\ref{sec:Dis_sim_vs_exp}.

\subsection{Effects of Atom--Surface Coupling Strength}\label{sec:result_coupling}

In the simulations, we set the coupling parameter
to $\eta = 3$~eV. The present framework allows us to vary $\eta$ systematically, providing mechanistic insight into how atom--surface coupling influences scattering dynamics. In earlier work,\cite{dan23b} we showed that coupling strength modulates the degree of adiabaticity: weak coupling favors nonadiabatic transitions, while strong coupling yields more adiabatic behavior, where the electronic friction theory becomes applicable. Here, we extend this analysis to H-atom scattering from a semiconductor surface.  

Figure~\ref{fig:Pop_eta} shows the time evolution of the neutral-state population $P_0(t)$ for different atom–-surface coupling strengths. As $\eta$ increases, a larger fraction of the wavepacket undergoes electron transfer to the surface, reflected by the reduced $P_0(t)$ around $t \approx 30$~fs. For $\eta > 2$~eV, the transfer probability saturates, with nearly complete electronic excitation during the collision. Stronger coupling also accelerates both the decay and recovery of $P_0(t)$, reflecting faster electron transfer. After scattering, the larger final value of $P_0(t)$ at higher $\eta$ indicates that more population returns to the neutral state and scatters back, while less remains trapped in the $|a\rangle$ state.  

\begin{figure}[h]
\centering
\includegraphics[width=0.8\linewidth]{Fig_pop0_eta.eps}
\caption{Time evolution of the H-atom population in the neutral state $\ket{0}$ for different coupling strengths $\eta$. The incident energy is $E_{\rm in}=1.92$~eV. All other parameters are identical to those in Table~\ref{Tab:para_all}.}
\label{fig:Pop_eta}
\end{figure}

Figure~\ref{fig:PE_eta} presents the projected kinetic energy distributions of the H atom in the neutral state at $t = 154.8$~fs for various $\eta$. 
At weak coupling, the elastic scattering channel is relatively pronounced, and its relative intensity decreases with increasing $\eta$, indicating a higher probability of electronic transitions, consistent with the population dynamics in Figure~\ref{fig:Pop_eta}.
Despite the change in intensity, the elastic peak remains centered near $E_{\rm in}$ and retains its initial shape for all coupling strengths.
This indicates that the elastic channel corresponds to scattering trajectories that are largely unaffected by the surface electronic bath.

The energy-loss peak also depends strongly on $\eta$: for weak coupling, the peak is broader, decreases in intensity, shifts significantly toward lower energies, and exhibits fine structure. 
These fine structures originate from discrete bound states of the excited potential $U_a(x)$ and higher-order processes such as multiple electron transfers. 
Specifically, when the H atom is excited to the $|a\rangle$ state, the bound states of $U_a(x)$ are discretely spaced rather than forming a continuum. 
These distinct excitation pathways can modulate the scattering dynamics, thereby shaping the observed structure in the energy distribution. 
Such fine structures are gradually washed out as $\eta$ increases.
% {\color{blue}
% These discrete excitation pathways modulate the scattering dynamics and are each broadened on the scale of the hybridization function’s bandwidth in Eq.~(\ref{Eq:level_width_func}), thereby shaping the observed structure in the energy distribution. 
% Such fine structures are gradually washed out as $\eta$ or the bandwidth (see Supporting Information) increases.}

\begin{figure}[h]
\centering
\includegraphics[width=0.8\linewidth]{Fig_PE_eta.eps}
\caption{Projected kinetic energy distributions of H atoms in the neutral state $\ket{0}$ at $t=154.8$~fs for various coupling strengths $\eta$. All other parameters are identical to those in Table~\ref{Tab:para_all}.}
\label{fig:PE_eta}
\end{figure}

The narrowing of the energy-loss peak and its shift toward higher kinetic energies with increasing $\eta$ indicate enhanced adiabaticity of the dynamics.
However, unlike H-atom scattering on metal surfaces, where the high-energy edge of the distribution typically approaches the initial kinetic energy, here the high-energy edge of the energy-loss peak remains around $1.5$~eV, roughly one band gap below the initial kinetic energy.
Thus, while strong coupling drives the dynamics toward adiabatic behavior, nonadiabatic effects remain intrinsic due to the semiconductor band structure. This persistent band-gap-related energy loss explains why the electronic friction theory fails to capture scattering on semiconductor surfaces.

\subsection{Effects of Incident Energy}

We now examine how the incident kinetic energy $E_{\rm in}$ influences the scattering outcome. Figure~\ref{fig:PE_Ein} shows the projected kinetic energy distributions of the H atom in the neutral state after scattering at different incident energies. Panel (a) corresponds to $\eta=1$ eV, while panel (b) corresponds to $\eta=3$ eV. 
The corresponding times chosen for $E_{in} = 0.37$, 0.99, 1.92, and 6.17 eV are $t = 164.5$, 193.5, 154.8, and 91.9 fs, respectively. 
For $E_{in}=0.37$ eV, a smaller box in the range $-30$ to $0.2$ a.u. was employed. In all cases, the outgoing wave packets have propagated well beyond the scattering region.

\begin{figure}[h!]
\centering
\includegraphics[width=0.5\linewidth]{Fig_PE_dfEin_eta13.eps}
\caption{Projected kinetic energy distributions of the H atom in the $|0\rangle$ state after scattering at different incident energies $E_{\rm in}$. (a) $\eta = 1$~eV. (b) $\eta = 3$~eV. 
For clarity, the distribution at $E_{in}=0.37$~eV in panel (b) is scaled by a factor of 3 due to its small magnitude.
All distributions were evaluated at sufficiently long times, when the outgoing wavepacket had moved far from the scattering region. 
All other parameters are identical to those listed in Table~\ref{Tab:para_all}.}
\label{fig:PE_Ein}
\end{figure}

For weak coupling ($\eta=1$~eV), the elastic peak persists at all incident energies. The relative weight of the energy-loss channel increases as $E_{\rm in}$ increases. At $E_{\rm in}=0.37$~eV (below the band gap), only elastic scattering is observed, with no energy-loss channel. At $E_{\rm in}=0.99$~eV, an energy loss feature appears below $E \approx 0.5$~eV, consistent with the Ge band gap. 
At $E_{\rm in} = 1.92$ eV, the energy-loss channel becomes more pronounced compared to the $E_{\rm in} = 0.99$ eV case. In both the $E_{\rm in} = 0.99$ and 1.92 eV cases, a substantial near-zero-energy component is present, originating from wavepacket components with kinetic energies insufficient to escape from the surface.
At the highest incident energy of 6.17 eV, the distribution in the energy-loss channel rises sharply from $E \approx 5.6$~eV toward lower energies, reaches a maximum, and then gradually decreases. Integration over the energy-loss channel shows that it contains an even larger contribution than that for $E_{\rm in} = 1.92$ eV. 

For strong coupling ($\eta=3$~eV), the elastic scattering peak is significantly suppressed across all incident energies. 
Note that the distribution at $E_{\rm in} = 0.37$~eV in panel (b) is multiplied by a factor of three for visibility.
The energy-loss peak appears only when $E_{\rm in}$ exceeds the band gap, and integration over the distributions confirms that the population in the energy-loss channel continues to grow with increasing $E_{\rm in}$. Consistent with the discussion in the previous subsection on the effect of atom–surface coupling strength, the profile of the energy-loss channel undergoes substantial changes with increasing $\eta$ for different $E_{in}$: the peak becomes smoother, shifts toward higher kinetic energies, and the low-energy component is reduced.

\subsection{Isotope Effect}

We now investigate the effect of isotope substitution on the scattering dynamics. Figure~\ref{fig:PE_isotope} shows the projected kinetic energy distributions $P_0^{\mathrm{pj}}(E)$ for H and D atoms at two incident energies ($E_{\rm in} = 1.92$ and 6.17~eV) and for both weak ($\eta = 1$~eV) and strong ($\eta = 3$~eV) atom–surface coupling. The left and right panels correspond to $E_{\rm in} = 1.92$ and 6.17~eV, respectively, while the top and bottom panels correspond to $\eta = 1$ and 3~eV, respectively. For D-atom simulations, all parameters are kept identical to those for H, except that the mass of D is set to be twice that of H. 

For reference, we also include the kinetic energy distribution of D at $E_{\rm in} = 1.87$~eV, which corresponds to the experimentally reported incident energy for D.\cite{kruger24}

The distributions are evaluated after the outgoing wavepacket has propagated well beyond the scattering region: $t=154.8$~fs for H and $t=222.5$~fs for D at $E_{\rm in} = 1.92$~eV; $t=91.9$~fs for H and $t=135.5$~fs for D at $E_{\rm in} = 6.17$~eV. 
For D at $E_{\rm in} = 1.87$~eV, the distributions are evaluated at $t=232.2$~fs.

\begin{figure}[h!]
\centering
\includegraphics[width=0.8\linewidth]{Fig_isotope_eff.eps}
\caption{Projected kinetic energy distributions in the $|0\rangle$ state from scattering simulations of H and D atoms at two incident energies and two coupling strengths. Panels (a)--(d) correspond to $(E_{\rm in}, \eta) =$ (1.92~eV, 1~eV), (6.17~eV, 1~eV), (1.92~eV, 3~eV), and (6.17~eV, 3~eV), respectively. 
For reference, in panels (a) and (c) we also include the distribution of D at 
$E_{\rm in}=1.87$~eV, shown as green dashed curves.
Distributions are evaluated when the outgoing wavepacket has moved far away from the scattering region for each case.}
\label{fig:PE_isotope}
\end{figure}

For weak coupling ($\eta=1$~eV), replacing H with D reduces the elastic scattering peak, enhances the energy-loss channel, and shifts its peak toward higher kinetic energies. For strong coupling ($\eta=3$~eV), where the elastic peak is already small, isotope substitution primarily narrows the energy-loss peak, reduces the low-energy component, and shifts the peak to higher energies. Comparison with Figure~\ref{fig:PE_eta} indicates that, for different $E_{\rm in}$ and $\eta$, replacing H with D consistently produces effects similar to increasing the atom--surface coupling strength.

This trend is physically intuitive. At the same incident energy, the heavier D atom has a lower velocity, thereby increasing the adiabaticity of the scattering process. 
This behavior can be qualitatively understood using the Landau--Zener formula for nonadiabatic transitions.\cite{landau32,zener32,nitzan06} According to this formula, the probability of a diabatic transition contains an exponential term with a factor $|V_{ab}|^2/v$, where $V_{ab}$ is the nonadiabatic coupling between diabatic states $|a\rangle$ and $|b\rangle$, and $v$ is the velocity. A decrease in $v$ thus has the same effect as an increase in $|V_{ab}|^2$. In our model, the coupling strength $\eta$ plays a role analogous to $|V_{ab}|^2$, so the isotope effect observed when replacing H with D mimics the effect of increasing $\eta$.

Experimentally, the D-atom scattering used for comparison with H at $E_{\rm in}=1.92$~eV was performed at a slightly lower incident energy of $E_{\rm in}=1.87$~eV. As indicated in Figure~\ref{fig:PE_isotope}(a) and (c), reducing the incident energy from 1.92 eV to 1.87 eV leads to a small shift of the peak toward lower energies.
A detailed comparison of the simulated isotope effect with experimental observations will be discussed in Sec.~\ref{sec:isotope_vs_exp}.

\section{Discussion}\label{sec:Discussion}

In this work, we present simulations of H-atom scattering on the Ge(111) surface. Our results show qualitative agreement with recent experimental findings, \cite{kruger23,kruger24} while certain discrepancies remain. Since our calculations employ the HEOM approach to obtain numerically exact solutions within the Newns–Anderson model framework, these deviations are likely due to intrinsic limitations of the model. This section is devoted to a detailed comparison between our theoretical predictions and the recent experimental observations.

\subsection{H-Atom Scattering: Simulation and Experiment}\label{sec:Dis_sim_vs_exp}

Our calculations predict a bimodal structure in the kinetic energy distribution of the scattered H atom, although the elastic scattering channel in this bimodal structure becomes less pronounced under strong coupling. This behavior originates from the gapped band structure of the semiconductor surface, indicating that the HEOM method provides a reliable nonperturbative description of H-atom scattering on semiconductor surfaces. Since the employed Newns--Anderson model accounts solely for interactions with surface electronic states and neglects surface phonons, this further supports that the bimodal structure arises from valence-to-conduction band excitations, producing high-energy electron–hole pairs.

Experiments have shown that the energy-loss channel is strongly promoted by the incident translational energy.\cite{kruger23} Our results exhibit the same trend: for both $\eta = 1$~eV and $\eta = 3$~eV, the integrated population of the energy-loss channel in Figure~\ref{fig:PE_Ein} increases with increasing incident energy. Analysis of the integrated population further indicates that the outgoing population grows with incident energy, consistent with the experimental observation that sticking probability decreases as the incident energy increases.\cite{kruger23,kruger24}

However, despite capturing this qualitative trend, the simulations do not reproduce the experimentally observed quantitative branching ratio between the elastic and energy-loss channels. We attribute this discrepancy to two factors. First, our model focuses on scattering at the Ge(111) rest site, whereas the experimental branching ratios reflect statistical averages over multiple surface sites. Second, as shown in Figure~\ref{fig:PE_eta}, the atom--surface coupling parameter $\eta$ significantly influences the relative weights of the two channels. 

Then, what magnitude of $\eta$ best corresponds to the actual case? Based on the results in Figs.~\ref{fig:PE_eta} and \ref{fig:PE_Ein}, we identify $\eta=3$~eV (strong coupling) as most representative, as the resulting energy-loss channel closely reproduces both the peak position and overall lineshape observed experimentally.\cite{kruger23} 
By contrast, although the $\eta = 1$~eV case exhibits a more pronounced bimodal structure, this feature cannot be taken as evidence that $\eta = 1$~eV is realistic, since its branch ratio also deviates significantly from experiment. 
Moreover, the branch ratio is influenced by factors beyond just $\eta$. For example, a larger bandwidth can make the elastic peak in the $\eta = 1$~eV case resemble that in the $\eta = 3$~eV case (see Supporting Information). Experimentally, the scattering results correspond to an average over different surface impact sites, so the branch ratio at a single site does not necessarily reflect the overall behavior. 
In particular, for the energy-loss channel, the $\eta = 1$~eV results differ substantially from the experimental observations, as the excess low-energy contributions lead to a marked shift in the peak position and an altered overall lineshape. The $\eta = 3$~eV value is also aligned with previous HEOM simulations of H scattering on Au(111),\cite{dan23b} suggesting a relatively consistent atom--surface coupling across different surfaces. 

To further validate this assessment, Figure~\ref{fig:PEloss_vs_Exp} compares the simulated energy-loss distributions at $\eta = 3$~eV for different incident energies with the experimental data from Fig.~1 of Ref.~\citenum{kruger23} 
(see the Supporting Information for a version of Figure~\ref{fig:PEloss_vs_Exp} that also includes the $\eta = 1$~eV results).
Overall, our simulations reproduce the key features of the experimental energy-loss channel. 
In particular, all curves correctly show that the minimum energy loss corresponds to the gap energy.
Some deviations remain: at $E_{in} = 0.99$~eV, the simulation overestimates the high energy-loss intensity due to contributions from low-energy distribution near $E \to 0$. This effect diminishes at higher incident energies as these components become less significant. At $E_{in} = 1.92$~eV, the simulated distribution nearly coincides with the experiment. For $E_{in} = 6.17$~eV, the results slightly overestimate the low-energy-loss intensity, whereas beyond the peak the distribution closely matches the experimental data.
These deviations may reflect limitations of our model PES and parameters, yet the overall agreement demonstrates that our simulations capture the essential mechanisms of the energy-loss process.
This agreement should nonetheless be interpreted with caution, as the one-dimensional model represents only rest-atom scattering and does not include other surface sites.

\begin{figure}[h!]
\centering
\includegraphics[width=0.8\linewidth]{Fig_dfEinEloss_vsExp.eps}
\caption{Translational energy-loss distributions for different incident energies. HEOM simulations ($\eta = 3$~eV, from Figure~\ref{fig:PE_Ein}; left y-axis) are plotted as a function of energy loss and compared with experimental data (normalized flux, right y-axis).
Experimental data adapted from Krüger, K. et al., Nature Chemistry 15, 326–331 (2023); licensed under a Creative Commons Attribution (CC BY 4.0) license.}
\label{fig:PEloss_vs_Exp}
\end{figure}

Since nearly all of the scattered wavepacket populates the energy-loss channel for $\eta=3$~eV, we further speculate that the two channels observed experimentally originate from collisions at different surface sites. Specifically, collisions of H with the surface rest atoms predominantly result in scattering through the energy-loss channel, while the elastic scattering channel mainly arises from collisions with other surface sites. Furthermore, the experimentally observed dependence of the branching ratio on $E_{\rm in}$ likely reflects the fact that changing the incident energy alters the relative probabilities of collisions at different surface sites. This finding provides direct support for the proposed site-specific scattering mechanism.\cite{kruger24,zhu24}

The agreement between our HEOM results and the experimental energy-loss channel is indeed surprising, given that our model uses simplified diabatic PESs fitted only to the rest-atom site and a hybridization function that does not capture the fine structure of the semiconductor band dispersion. This nevertheless suggests that the Newns–Anderson Hamiltonian captures the essential physics of H–Ge(111) scattering and that the energy-loss channel is indeed dominated by electron–hole pair excitation. Moreover, as discussed in Sec.~\ref{sec:result_coupling}, the dynamics exhibit an underlying adiabatic character in the realistic (stronger-coupling) regime, which reduces the sensitivity to the detailed construction of the diabatic PESs. In this regime, even though electronic friction theory does not apply, certain approximate conditions may still hold. 
The fact that a simple hybridization function already yields a realistic energy-loss distribution further suggests that the dynamics depend only weakly on the fine features of the semiconductor band structure, and that its overall influence, much like the friction coefficient in electronic friction theory, may likewise be effectively captured by just a small set of parameters.

\subsection{Isotope Effects: Simulation and Experiment}\label{sec:isotope_vs_exp}

% Recently, isotope effect measurements\cite{kruger24} show that for the adiabatic channel (or elastic scattering channel in this work), deuterium exhibits a larger energy loss than H, whereas the energy loss channel shows almost no isotope effect. 
% Our simulation results do not capture the broadening of the elastic peak with increasing atomic mass. But it is noted that the elastic channel is not our main focus, as it can already be reliably described by adiabatic MD simulations. Moreover, the Newns–Anderson model used here does not include coupling to surface phonons and is therefore not intended to reproduce adiabatic dynamics involving phonon interactions quantitatively.

Recently, isotope effect measurements\cite{kruger24} show that for the adiabatic channel (or elastic scattering channel in this work), deuterium exhibits a larger energy loss than H, whereas the energy loss channel shows almost no isotope effect. 
At first glance, this may appear to contradict our simulations of the isotope effect. However, we will show that the theoretical results do not contradict the experimental findings and instead provide a more detailed understanding.

Regarding the elastic peak, our simulation results do not capture the broadening of the elastic peak with increasing atomic mass. 
This is not a concern, as the elastic channel is not the main focus of this study and can already be reliably described by adiabatic MD simulations. Furthermore, the Newns–Anderson model used here does not include coupling to surface phonons and is therefore not intended to reproduce adiabatic dynamics involving phonon interactions quantitatively.

For the more relevant energy-loss channel, 
which is strongly influenced by atom--surface interactions and directly linked to surface electronic properties, our simulations offer complementary insights and reveal new aspects of isotope effects.
Although the experiment reports “almost no isotope effect” in this channel, a closer inspection of the energy-loss spectra (Figure~2 of
Ref.~\citenum{kruger24}) reveals a slight but systematic difference between the H and D energy-loss peaks, with the D peak shifted toward smaller energy loss. 
This subtle trend is consistent with our prediction in Figure~\ref{fig:PE_isotope}, where increasing the atomic mass produces an effect analogous to a modest increase in the atom–surface coupling strength, shifting the deuterium peak toward higher energies (i.e., smaller energy loss). On the other hand, in the experiment the deuterium data were taken at a slightly lower incident energy ($E_{\rm in}=1.87$ eV), which shifts the peak toward lower energies relative to the $E_{\rm in}=1.92$ eV condition used for H. These two effects partially cancel each other, resulting in the very weak isotope dependence observed experimentally.

We also acknowledge that certain discrepancies remain between theory and experiment. In particular, the peak heights and widths of H and D in Figure~\ref{fig:PE_isotope} differ more visibly than in the measured spectra. Part of this difference arises because the experimental signals are normalized and represent a statistical average over all adsorption sites, which suppresses features associated with the rest site. Another possible source is the limitation of our model parameters, such as the choice of diabatic potential energy surfaces.

\section{Conclusions}\label{sec:Conclusion}

We have employed the numerically exact hierarchical equations of motion (HEOM) method to simulate H-atom scattering from the Ge(111) surface.  The one-dimensional PES for the H atom colliding with the rest site of Ge(111) was obtained from DFT calculations and fitted to the Newns–Anderson model to determine the model parameters. The simulations reveal two distinct scattering channels: an elastic channel without electron transfer and an energy-loss channel involving electron--hole excitations across the semiconductor band gap. These two channels produce a bimodal kinetic energy distribution of the scattered H atom, with peak separation consistent with the band gap.

Systematic variations of the atom--surface coupling strength, incident energy, and isotope substitution reveal clear mechanistic insights. Increasing the coupling strength drives scattering predominantly into the energy-loss channel, reduces both the energy loss and the broadening of the outgoing kinetic energy distribution, and enhances the adiabaticity of the dynamics.  Higher incident energies promote the energy-loss channel, generating the bimodal structure. Isotope substitution (H $\to$ D) produces a similar impact as increasing the atom--surface coupling strength. 

Importantly, under strong atom--surface coupling, the overall shape and peak position of the simulated energy-loss peak closely match experimental observations while providing additional microscopic insight. HEOM captures high-energy electron–hole excitations in the energy-loss channel that cannot be described by electronic friction theory and other perturbative approaches,\cite{kruger23} addressing a key limitation of existing BOA-based methods. 
However, under this strong coupling, the elastic scattering channel is underestimated, supporting the experimentally proposed site-specific scattering mechanism, in which elastic scattering from other sites also contributes. Furthermore, while the experimental energy-loss peak appears relatively insensitive to isotope substitution, our results reveal a systematic shift of the deuterium peak toward smaller energy loss, highlighting the ability of HEOM simulations to uncover subtle quantum effects masked in ensemble-averaged measurements.

The fully quantum HEOM approach thus provides a benchmark for understanding H–semiconductor scattering at a fundamental level. 
Due to the substantial computational cost, we have employed a simplified Newns–Anderson model. 
In the future, a more complete model could include all four H electronic states, incorporating Coulomb repulsion and potentially revealing additional physical phenomena, while the present study provides a solid foundation for such extensions. Moreover, the model could also be extended to include both surface electronic and phonon baths by employing HEOM for hybrid fermionic–bosonic environments.\cite{ke22}
Extending these simulations to fully realistic surface environments will additionally require the development of efficient approximate methods.

Nonetheless, our fully quantum results offer valuable mechanistic insights and guidance for future modeling. As demonstrated, realistic scattering conditions correspond to a strong atom–surface coupling regime characterized by relatively adiabatic dynamics, albeit with characteristic band-gap-controlled energy loss. Incorporating such band-gap effects into extensions of electronic friction theory and related frameworks represents a promising avenue for future research.

% Beyond reproducing specific experimental features, our results highlight how semiconductor band gaps impose a fundamental lower bound on energy loss in surface scattering. This band-gap-controlled nonadiabaticity has direct implications for heterogeneous catalysis and photocatalysis, where hot electron and hole excitations can drive chemical transformations. The ability of HEOM to capture these effects nonperturbatively opens the door to predictive modeling of energy dissipation and charge-transfer processes at realistic semiconductor interfaces. Such insights are essential for optimizing catalytic selectivity, understanding energy conversion efficiency, and ultimately guiding the design of advanced catalytic and energy materials.  

\section*{Supplementary Material}
See the supplementary material for additional HEOM simulation results and supporting figures.

\section*{Acknowledgments}

V.~S.~B. acknowledges support from the U.S. Army Research Office under Award W911NF-21-1-0337.  
X.~D. thanks Dr.~L. Zhu for helpful discussions regarding the computational results reported in their work, and Dr.~O. Bünermann for kindly sharing the experimental data, and the Yale Center for Research Computing for a generous allocation of HPC time. 
Q.~S. acknowledges support from the National Natural Science Foundation of China (Grant No.~22433006).

\section*{Availability of data}
The data that support the findings of this study are available from the corresponding author upon reasonable request.

%%%%%%%%%%%%%%%%
\pagebreak

\begin{table}[h!]\label{Tab:para_all}
\begin{tabular}{c  c}
\toprule [2pt]
\parbox[c]{4cm}{\centering Parameter} & \parbox[c]{6cm}{\centering Value} \\
\midrule [1pt]
$\eta$ & $3$~eV \\
$\epsilon_0$ & $0.0$~eV   \\ 
$\gamma$ & $1.5$~eV   \\
$E_B$ & $0.49$~eV   \\
$\delta$ & $0.02$~eV   \\
$\mu$ & $0.0$~eV   \\
$M$ & 1836.013~au \\
$T$ & 300~K   \\
$A_0$ & $0.19$~eV \\
$C_0$ & $1.2$~au \\
$x_0$ & $-3.8$~au \\
$A_1$ & $11.81$~eV \\
$B_1$ & $8.8$~eV \\
$C_1$ & $0.32$~au \\
$D_1$ & $0.94$~au \\
$x_1$ & $-0.898$~au \\
$c_g$ & $1.1$~au   \\
$x_g$ & $-2.8$~au  \\
\bottomrule [2pt]
\end{tabular}
\caption{Model parameters for the Newns-Anderson model.}
\end{table}
%%%%%%%%%%%%%%%%%% 

\bibliography{./quantum}

\end{document}

% \title{Nonadiabatic H-Atom Scattering Channels on Ge(111) Captured by MPS-HEOM}
\title{Supporting Information: Nonadiabatic H-Atom Scattering Channels on Ge(111) Elucidated by the Hierarchical Equations of Motion}
\author{Xiaohan Dan}
\affiliation{Department of Chemistry, Yale University, New Haven, CT 06520,
USA}
\author{Zhuoran Long}
\affiliation{Department of Chemistry, Yale University, New Haven, CT 06520, USA}
\author{Tianyin Qiu}
\affiliation{Department of Chemistry, Yale University, New Haven, CT 06520, USA}
\author{Jan Paul Menzel}
\affiliation{Department of Chemistry, Yale University, New Haven, CT 06520, USA}

\author{Qiang Shi}\email{qshi@iccas.ac.cn}
\affiliation{Beijing National Laboratory for Molecular Sciences, State Key Laboratory for Structural Chemistry of Unstable and Stable Species, CAS Research/Education Center for Excellence in Molecular Sciences, Institute of Chemistry, Chinese Academy of Sciences, Zhongguancun, Beijing 100190, China, and University of Chinese Academy of Sciences, Beijing 100049, China}

\author{Victor S. Batista}\email{victor.batista@yale.edu}
\affiliation{Department of Chemistry, Yale University, New Haven, CT 06520, USA}
\affiliation{Yale Quantum Institute, Yale University, New Haven, CT 06511, USA}

\maketitle

%To mark blue Since the whole SI is new 
% \color{blue}

\section{Projected and Unprojected Distributions}
Here we illustrate the effect of applying the projection operator defined in Sec.~III~C of the main text to exclude near-surface contributions. 
Figure~\ref{fig:Px_befaftpj} shows the spatial probability distributions of the H atom in the $|0\rangle$ state at the outgoing time ($t = 154.8$ fs), before and after applying the projection, for $\eta = 1$ and 3 eV. The incident energy is $E_{\mathrm{in}} = 1.92$ eV. 
Figure~\ref{fig:PE_befaftpj} shows the corresponding projected kinetic energy distributions under the same conditions.

\begin{figure}[H]
\begin{center}
\includegraphics[width=0.7\linewidth]{Fig_Px6400_projbefvsaft_eta13.eps}
\end{center}
\caption{
Spatial probability distributions of the H atom in the $|0\rangle$ state at the outgoing time ($t = 154.8$~fs), shown before and after applying the projection operator defined in Sec.~III~C of the main text to remove near-surface trapped components. Results are given for $\eta = 1$ and 3 eV. 
All other parameters follow Table I of the main text.
The incident energy is $E_{\mathrm{in}} = 1.92$ eV.
}
\label{fig:Px_befaftpj}
\end{figure}

\begin{figure}[H]
\begin{center}
\includegraphics[width=0.7\linewidth]{Fig_PE6400_projbefvsaft_eta13.eps}
\end{center}
\caption{Kinetic energy distributions of the emitted H atom before and after applying the projection operator, under the same conditions as in Figure~\ref{fig:Px_befaftpj}. The removal of near-surface trapped components only affects the low-energy region near $E \to 0$.}
\label{fig:PE_befaftpj}
\end{figure}

\section{Influence of the Bandwidth on the Dynamics}
In the main text, we adopted a hybridization function of the Lorentzian form with an explicit band gap,
\begin{equation}\label{Eq:level_width_func}
\Gamma(\epsilon) = 
\frac{\eta \gamma^2}{(\epsilon-\epsilon_0)^2 + \gamma^2}
\left[ 1 - \frac{1}{1+e^{(\epsilon-E_B/2)/\delta}}
          \frac{1}{1+e^{-(\epsilon+E_B/2)/\delta}} \right],
\end{equation}
with a bandwidth $\gamma = 1.5$~eV. Here, we examine how varying the bandwidth affects the dynamical results and confirm that the final energy-loss distributions---especially for the more realistic coupling strength $\eta=3$~eV---are largely insensitive to this choice.

We tested three representative bandwidths, $1.5$ eV, $3$ eV, and $5$ eV, whose hybridization functions are shown in Figure~\ref{fig:Jome_dfwidth}. The largest bandwidth already spans over 10 eV, well beyond the energetically relevant window for the scattering dynamics. Figure~\ref{fig:PE_dfbandwidth} shows the kinetic energy distributions of H atoms after scattering ($t = 154.8$~fs). The incident energy is $E_{in}=1.92$~eV, and all simulation parameters other than the bandwidth and $\eta$ follow Table I of the main text.

\begin{figure}[H]
\begin{center}
\includegraphics[width=0.5\linewidth]{Fig_Jome_dfwidth.eps}
\end{center}
\caption{Hybridization functions for bandwidths $\gamma = 1.5$~eV, $3$~eV, and $5$~eV. All other parameters follow Table I of the main text.}
\label{fig:Jome_dfwidth}
\end{figure}

Figure~\ref{fig:PE_dfbandwidth} shows the kinetic energy distributions of H atoms after scattering for two incident energies, $E_{in}=1.92$~eV and $E_{in}=6.17$~eV, and for two coupling strengths, $\eta=1$~eV and $\eta=3$~eV. The distributions are evaluated at $t = 154.8$~fs for $E_{in}=1.92$~eV and $t = 91.9$~fs for $E_{in}=6.17$~eV.

\begin{figure}[H]
\begin{center}
\includegraphics[width=0.8\linewidth]{Fig_PE_dfbandwidth.eps}
\end{center}
\caption{Projected kinetic energy distributions of H atoms in the $|0\rangle$ state after scattering, shown for bandwidths $\gamma = 1.5$~eV, $3$~eV, and $5$~eV at two incident energies and two coupling strengths. Panels (a)--(d) correspond to $(E_{\rm in}, \eta) =$ (1.92~eV, 1~eV), (6.17~eV, 1~eV), (1.92~eV, 3~eV), and (6.17~eV, 3~eV), respectively. 
Distributions are evaluated once the outgoing wavepacket has propagated sufficiently far from the scattering region. All simulation parameters other than the bandwidth and $\eta$ follow Table I in the main text.}
\label{fig:PE_dfbandwidth}
\end{figure}

For the weak atom–surface coupling $\eta=1$~eV in Figure~\ref{fig:PE_dfbandwidth}(a) and (b), increasing the bandwidth consistently reduces the elastic channel. 
A broader band enables the projectile to couple to a larger manifold of surface states and thereby increases the probability of electron–hole pair excitations during the collision. 
For the lower incident energy $E_{in}=1.92$~eV [panel (a)], when the bandwidth is increased to $\gamma=3$~eV, the electron-transfer probability between the H atom and the surface approaches saturation. Further increasing the bandwidth produces almost no additional reduction of the elastic channel, leaving a small residual elastic component.

For the energy-loss channel, increasing the bandwidth causes a slight increase in intensity, consistent with the reduction in the elastic channel. In addition, increasing $\gamma$ from 1.5 to 3~eV slightly broadens the energy-loss peak, leading to more distribution in the low-energy region.
Beyond $\gamma = 3$~eV, the bandwidth effects become negligible, and increasing the bandwidth to 5~eV results in almost no further change in the distribution.

For strong coupling ($\eta = 3$ eV) in Figure~\ref{fig:PE_dfbandwidth}(c) and (d), the electron-transfer probability is already close to saturation, and the elastic peak is therefore very small. In this regime, increasing the bandwidth produces only minor changes in the results. For $E_{in}=1.92$ eV, the kinetic energy distribution is essentially insensitive to the bandwidth. For $E_{in}=6.17$ eV, bandwidth effects are slightly more pronounced: increasing $\gamma$ from 1.5 eV to 3 eV leads to a modest broadening of the main energy-loss peak, whereas the change from 3 eV to 5 eV is negligible. These observations indicate that, under strong coupling, increasing the bandwidth drives the dynamics toward the wide-band limit, and the wide-band limit is expected to yield results comparable to those shown here.

Since the main tests correspond to the strong-coupling regime, all conclusions drawn in the main text using $\gamma = 1.5$~eV remain valid. The small changes observed at $E_{\rm in} = 6.17$~eV do not affect the main results or the good agreement with experiment shown in Figure 10 of the main text. Furthermore, because $P_0^{\rm pj}(E \to 0)$ slightly decreases for larger $\gamma$, the comparison with experimental data is even marginally improved.

% \begin{figure}[H]
% \begin{center}
% \includegraphics[width=0.7\linewidth]{Fig_PE6400_dfbandwidth_eta13.eps}
% \end{center}
% \caption{Projected kinetic-energy distributions of H atoms after scattering ($t = 154.8$~fs) for bandwidths $\gamma = 1.5$~eV, $3$~eV, shown for (a) weak coupling $\eta = 1$~eV and (b) strong coupling $\eta = 3$~eV. The incident energy is $E_{in}=1.92$~eV, and all other simulation parameters follow Table I of the main text.}
% \label{fig:PE_dfbandwidth_eta13}
% \end{figure}

% \begin{figure}[H]
% \begin{center}
% \includegraphics[width=0.7\linewidth]{Fig_PE_dfbandwidth_Ein6p17_eta13.eps}
% \end{center}
% \caption{Projected kinetic-energy distributions of H atoms after scattering ($t = 154.8$~fs) for bandwidths $\gamma = 1.5$~eV, $3$~eV, shown for (a) weak coupling $\eta = 1$~eV and (b) strong coupling $\eta = 3$~eV. The incident energy is $E_{in}=6.17$~eV, and all other simulation parameters follow Table I of the main text.}
% \label{fig:PE_dfbandwidth_eta13_Ein6p17}
% \end{figure}

\section{Effect of the Initial Momentum Distribution}

In scattering experiments, the kinetic energy distributions of atomic hydrogen beams typically have widths of only a few tens of meV.\cite{bunermann15b} 
In contrast, the initial kinetic-energy spread used in our simulations is on the order of several hundred meV. 
Nevertheless, according to the position–momentum uncertainty principle, a narrower initial kinetic-energy distribution necessarily corresponds to a broader initial spatial distribution, which in turn requires a larger simulation box and longer propagation times. Because our calculations are fully quantum mechanical, we must acknowledge that they remain constrained by computational resources, reproducing near-macroscopic experimental conditions ($\mu s$ time scales and $mm$ spatial extents\cite{bunermann15b}) is not feasible.

To demonstrate that this broader initial distribution does not significantly affect the resulting dynamics, we performed HEOM simulations using different initial kinetic-energy widths while fixing the mean incident energy at $E_{\mathrm{in}} = 1.92$ eV. All other parameters were kept unchanged, and only the width $\sigma_x$ of the initial Gaussian wavepacket was varied. The corresponding initial kinetic-energy distributions are shown in Figure~\ref{fig:PE_dfsigma_init}; note that a smaller $\sigma_x$ leads to a broader distribution in energy space.

\begin{figure}[H]
\begin{center}
\includegraphics[width=0.5\linewidth]{Fig_PE_dfsigma_init.eps}
\end{center}
\caption{Initial kinetic energy distributions corresponding to different widths $\sigma_x$ of the Gaussian wavepacket, all centered at $E_{\mathrm{in}} = 1.92$~eV.}
\label{fig:PE_dfsigma_init}
\end{figure}

Figure~\ref{fig:PE_dfsigma_final} presents the projected kinetic energy distributions at $t = 154.8$ fs for $\eta = 1$ eV (panel a) and $\eta = 3$ eV (panel b). The elastic-scattering peak—originating from trajectories that do not undergo electron transfer with the surface—directly reflects the characteristics of the initial kinetic-energy distribution. Consequently, reducing the initial energy spread leads to a corresponding narrowing of this elastic peak. 
By contrast, for the energy-loss channel, apart from a modest broadening when $\sigma_x$ increases from 0.5 to 1.0, further increases in $\sigma_x$ (i.e., further narrowing of the initial energy distribution) produce almost no change in the peak shape or width. 
These results indicate that the energy-loss peak is largely insensitive to the precise width of the initial kinetic-energy distribution. Therefore, the choice of $\sigma_x = 1.0$ used in the main text is justified, and we expect that the physical conclusions regarding nonadiabatic scattering (the energy-loss channel) would remain unchanged under narrower, experimentally realistic energy spreads.

\begin{figure}[H]
\begin{center}
\includegraphics[width=0.7\linewidth]{Fig_PE_dfsigma_final_e13.eps}
\end{center}
\caption{Projected kinetic energy distributions of the H atom at $t = 154.8$~fs for (a) $\eta = 1$~eV and (b) $\eta = 3$~eV, obtained using different initial wavepacket widths $\sigma_x$.
The mean incident energy is fixed at $E_{\mathrm{in}} = 1.92$~eV, and all other simulation parameters follow Table I of the main text. The width of the elastic-scattering channel follows the width of the initial kinetic-energy distribution, whereas the energy-loss channel remains largely insensitive to $\sigma_x$, aside from a noticeable broadening when $\sigma_x$ is increased from 0.5 to 1.0.}
\label{fig:PE_dfsigma_final}
\end{figure}

\section{Energy-Loss Distributions vs Experiment at Different Couplings}
Figure~\ref{fig:dfEloss_e13} compares the simulated energy-loss distributions at $\eta=1$~eV and $\eta=3$~eV for different incident energies with the experimental data. 
Same as in Figure~10 of the main text, except that we additionally include the $\eta = 1$~eV results for a more explicit comparison. Although the $\eta=1$~eV case exhibits a more pronounced bimodal structure, its energy-loss channel (the higher–energy-loss peak) deviates significantly from the experimental measurements. This comparison further supports our argument in the main text that $\eta=3$~eV corresponds more closely to the experimental situation.

\begin{figure}[H]
\begin{center}
\includegraphics[width=0.7\linewidth]{Fig_dfEinEloss_vsExp_eta13.eps}
\end{center}
\caption{Translational energy–loss distributions from HEOM simulations (for $\eta = 1$ and $3$~eV) at various incident energies. The distributions are plotted as a function of the energy loss and compared with the experimental normalized flux.
Experimental data adapted from Krüger, K. et al., Nature Chemistry 15, 326–331 (2023); licensed under a Creative Commons Attribution (CC BY 4.0) license.
}
\label{fig:dfEloss_e13}
\end{figure}

%back to black color
% \color{black}

% %%%%%%%%%%%%%%%%
% \pagebreak
\bibliography{./quantum}